# Kink-pair mechanism and low temperature flow-stress behaviour of strontium titanate single crystals

M. Castillo-Rodríguez and W. Sigle

*Max Planck Institute for Metals Research, StEM, Heisenbergstr. 3, 70569 Stuttgart, (Germany), castillo@mf.mpg.de, sigle@mf.mpg.de*

The mechanical behaviour of strontium titanate exhibits a remarkable behaviour at low temperature, in the so called regime A, where the flow stress experiences two different temperature dependences separated by a noticeably abrupt drop in between. The dislocation microstructure was investigated and, by making adequate use of the kink-pair model, we interpret this behaviour as a transition from the short- to the long-segment limit of kink-pair formation. The fit parameters are found to be physically sound.

Strontium titanate is one of the most important oxides whose mechanical behaviour has been investigated in a wide temperature range (46–1811 K) [1-6]. The high-temperature behaviour has been extensively studied and the ductile-to-brittle-to-ductile transition (DBDT) exhibited between 1000 and 1500 K can be explained by the climb dissociation of $a$<110> *edge* dislocation on a {110} plane [5]. However, below room temperature the flow stress increases in a non-monotonic manner with temperature [3] (Fig. 1). Down to 225 K the flow stress dependence with temperature is rather flat (Regime $A_1$), but between 225 and 200 K the flow stress increases by almost a factor of two (Regime $A_2$), followed by a strong further increase below 200 K (Regime $A_3$). A similar discontinuity is known from body-centered cubic (BCC) metals like Fe [7], Nb and Nb alloys [8], and Mo [9]. Since at low temperature the motion of screw dislocations is often controlled by a kink-pair mechanism [10], the flow-stress discontinuity was explained as a change of kink-pair plane [11] or kink height [12]. In this work we have investigated the dislocation microstructure of $SrTiO_3$ samples deformed from regime $A_1$ to $A_3$, observing that there is a slip preference along the <110> direction in regime $A_3$ which gradually disappears with increasing temperature and vanishes in regime $A_1$. Based on this preference, we have used a kink-pair model to fit our experimental data, considering a transition from the "short-segment limit" in regime $A_3$ to the "long-segment limit" in regime $A_1$. The physical parameter values provided by the best fit will be discussed and we show that they all are physically sound, which strongly supports this interpretation.

Quadrangular prisms of size (2.5 x 2.5 x 6) $mm^3$ of $SrTiO_3$ single crystals were deformed in compression parallel to the <001> direction with a constant displacement rate of $\dot{\varepsilon}$ ~ $10^{-4}$ $s^{-1}$ [3], activating the <110>{110} slip systems. Specimens for transmission electron microscopy (TEM) were prepared, following standard procedures, parallel to the most activated slip plane for weak-beam dark-field (WB-DF) imaging observations. The dislocation microstructure was investigated by WB-DF in a Philips CM200 microscope.

The temperature dependence of the critical resolved shear stress (CRSS) was analysed in terms of the kink-pair model. In the short-segment limit [13], the dislocation portion on which the kink pair nucleates is short enough to allow its expansion over the full dislocation length $L$. In this case the strain rate is given by:

$$\dot{\varepsilon} = \rho b L \upsilon \frac{\tau' b_1 h^2}{\alpha kT} \exp\left(-\frac{Q'_D + 2F'_k}{kT}\right) \exp\left(\frac{\left(\tau' b_1^3 h^3 \mu\right)^{1/2}}{(2\pi)^{1/2} kT}\right) \qquad (1)$$

where $\rho$ is the mobile dislocation density whose dependence on stress has been taken as $\rho = \left(\dfrac{\tau}{C\mu b_1}\right)^2$ with $C$ being a constant [14], $b$ the Burgers vector of the perfect dislocation and $b_1$ the partial Burgers vector. $\upsilon$ is the attempt frequency (typically the Debye frequency $\sim 10^{12}$-$10^{13}$ s$^{-1}$), $h$ the periodicity of the primary Peierls potential, $\mu$ the shear modulus ($\sim$112.5 GPa, [15]), $Q_D^{'}$ the height of the secondary Peierls barrier, $F_K^{'}$ the free energy of a single kink on a partial dislocation, $\gamma$ the stacking-fault energy, $k$ the Boltzmann constant, $T$ the absolute temperature and $\alpha$ a factor given by:

$$\alpha = \frac{1}{2}\left\{1+\left[1+\left(\frac{\lambda \tau' \mu b^3 h^3}{\pi K^2 T^2}\right)^{1/2}\right]^{1/2}\right\} \quad (2)$$

where $\lambda$ is a parameter equal to 2 or 4 for kink pair nucleation on the leading partial or simultaneously on both partials, respectively. $\tau'$ is the effective stress on the portion of the leading partial given by:

$$\tau' = \tau - \tau^{\gamma} \quad (3)$$

$\tau$ is the resolved shear stress applied to the specimen, and $\tau^{\gamma} = \dfrac{\pi \gamma^2 h}{\mu b_1^3}$ is the resolved applied stress required for the nucleation in the leading partial due to the required extra stacking-fault energy.

On the other hand, in the long-segment limit [13], kink pairs expand until they annihilate with other kinks from vicinal sources along the same partial dislocation line. Following the same procedure as in [13], but considering that nucleation takes place simultaneously in both partials, we obtain that the strain rate is given by:

$$\dot{\varepsilon} = \rho b a \upsilon \frac{\tau' b_1 h^2}{\alpha^{1/2} kT} \exp\left(-\frac{Q_D' + 2F_k'}{kT}\right) \exp\left(\frac{(\tau' b_1^3 h^3 \mu)^{1/2}}{(4\pi)^{1/2} kT}\right) \quad (4)$$

where $a$ is the periodicity of the secondary Peierls potential. It is worth emphasizing that in the limit $b_1 \rightarrow b$ and $\tau' \rightarrow \tau$, expressions (1) and (4) correspond to nucleation on an undissociated dislocation ~~or to correlated nucleation in both partials~~.

We have performed TEM observations in samples deformed at different temperatures from regime $A_1$ to regime $A_3$ (Fig. 2). First of all, it is worth emphasizing that the dislocation microstructure observed in micrographs does not exactly correspond to the one present just at the end of the compression test, since samples are warmed up or cooled down from the temperature of the test to room temperature. This means that the dislocation microstructure can undergo a recovery process, like it has been detected in NiO single crystals deformed below room temperature [16]. In this recovery process dislocations rearrange themselves into configurations of lower energy. Basically, dislocations tend to lie along close-packed directions in which the Peierls potential exhibits minima and dipoles tend to rotate to reach edge character which decreases their energy. As we show in this paper, the presence of a large amount of screw dipoles in samples strained in regime $A_3$ and bent dislocations in regime $A_1$ seems to indicate that these recovery processes are of minor importance for the final dislocation structure of the samples we have investigated.

At lowest temperatures (Regime A3) the dislocation microstructure is governed essentially by $a$<110> screw dislocations decorated by short edge segments. This is evidence that in this temperature range dislocation glide along the <110> direction is favored against the <100> direction. This means that during the dislocation movement glide along the <110> direction is easier and thus much faster than along the <100> direction, so that consequently dislocation lines are elongated along <110> directions, as it is experimentally observed. In the kink-pair model this means that for the

*a*<110> screw dislocations glide is controlled by kink-pair nucleation along the <100> direction. As temperature increases, in regime $A_1$ dislocations are strongly bent, indicating that the screw preference vanishes and the difference in glide along <100> and <110> is less significant than in regime $A_3$. This indicates that in regime $A_1$ kink-pair nucleation and expansion becomes easier than at lower temperatures, i.e. the probability per unit dislocation length for kink-pair nucleation increases with temperature and consequently the probability of a kink pair to be annihilated with another kink before expanding until the full dislocation length also increases. Therefore it is likely that in regime $A_3$ a kink-pair nucleation mechanism in the short-segment limit is operative whereas in regime $A_1$ the long-segment limit controls dislocation glide. Therefore the drop of the CRSS in regime $A_2$ could be due to a transition of the dislocation glide mechanism. The origin of this transition could be an increasing mobility of atoms in the temperature regimes $A_2$ and $A_1$, making atomic rearrangements easier. This could facilitate kink formation and justify a change from the short to the long segment limit mechanism.

We now use this model to fit the experimental temperature dependence of the CRSS. We have fixed some of the parameter values of the model, and the numerical optimization of the least square fit between the experimental CRSS and the theoretical prediction of equations (1) and (4) provides us with values of the other free parameters. A reliable fit is characterized by a small deviation from the experimental data and by physically sensible values of the free parameters.

Figure 1 shows the best fit of the kink-pair model to the experimental data. Since dislocations have mainly screw character, we have taken the lattice parameter for the periodicity of the primary Peierls potential $h$ = 0.3905 nm. For the dislocation length $L$ we have selected the mean distance between dislocations which is an approximation of the dislocation length. From TEM micrographs we estimated $L$ = 400 nm. For the magnitude of the Burgers vector we took $b_1$=0.552/2 nm, which is the Burgers vector of the two collinear partials of an *a*<110> dislocation. The value of the stacking-fault energy was taken from our recent measurement [17]: $\gamma$ = 340 ± 90 mJ/m$^2$. In regime A$_3$ *a*<110> dislocations are dissociated by glide as it has been observed [17] and we assume that kink-pair nucleation takes place individually in each partial. This assumption is supported by two experimental observations: (i) the relatively small stacking-fault energy in the {110} plane does not make simultaneous kink-pair formation on both partials necessary because the energy increase by the formation of one kink pair is moderate. (ii) Calculating the resolved applied stress required for a kink-pair nucleation in one partial from Eq. (3) we obtain a value of $\tau^\gamma$ ~ 55 MPa. Since in regime A$_3$ the resolved applied stress is above 100 MPa, i.e. higher than $\tau^\gamma$, separate nucleation in each partial is possible. Finally, the difficulty exhibited by dislocations to glide along the <100> direction could be another argument that kink-pair nucleation does not take place correlatively in both partials. The best fit in this regime A$_3$ provides us with a value for the dislocation density $\rho$ between 10$^{11}$ and 10$^{12}$ m$^{-2}$ which is typical for the dislocation density of a strained material. With regard to the energy we obtain that $q$(A$_3$) = $Q_D$'(A$_3$) + 2$F_k$'(A$_3$) = 1.1 ± 0.2 eV. We will later on analyse and compare this value with regime A$_1$.

In regime A$_1$ (temperature range 225–1000 K), we consider kink-pair nucleation on a perfect screw dislocation and we apply the long-segment limit. The first assumption is due to the resolved applied stress value which is between 65 and 45 MPa along the whole regime A$_1$, the mean value of which coincides well with $\tau^\gamma$ ~ 55 MPa. This means that now kink-pair nucleation will take place in both partials simultaneously, since otherwise almost all the resolved applied stress would be required to overcome the resolved stress coming from the extra stacking-fault energy of the leading partial. The long-segment limit is used because in this limit dislocations tend to become bent by kink pair annihilation. This type of bent dislocations is indeed observed in TEM micrographs. Here we also want to mention that we do not take into account climb dissociation of *a*<110> edge dislocations, since this mechanism seems to take place significantly only near the high-temperature end of regime $A_1$ and particularly in regimes B and C where it leads to the ductile-to-brittle-to-ductile transition (DBDT) [5]. Consequently, at the end of regime $A_1$ (above 1000 K) we expect the experimental data to deviate from our fit.

For the periodicity of the secondary Peierls potential *a* we took the length of the perfect *a*<110> Burgers vector $b = 0.552$ nm. The best fit provides us with a value for the dislocation density of about $\rho \sim 10^{12}$ m$^{-2}$, which is within typical values of strained materials. However, in this regime dislocations are bent and the preference of glide along <110> against <100> seems to have vanished. For the energy we obtain $q(A_1) = Q'(A_1) + 2F'_k(A_1) = 0.6 \pm 0.1$ eV which is much lower than in regime $A_3$ indicating that the kink-pair nucleation and migration is much easier than in regime $A_3$. This higher nucleation rate can explain the dislocation bending observed in samples deformed in this regime.

The activation energies in each regime can also be determined approximately from an Arrhenius-type relationship, like it has been reported previously [3], obtaining an activation enthalpy of about ~0.58 eV for regime $A_3$ and ~0.75 eV for regime $A_1$. Although the activation energy we obtain for regime $A_1$ coincides reasonably well, for regime $A_3$ the difference is large. The reason could come from our assumption, argued at the beginning of this work, that in regime $A_3$ kink pair nucleation takes place on the leading partial instead of the perfect dislocation, like it was considered in

[3]. In that case, eq. (1) would be modified and then would provide different values from the ones shown here for regime $A_3$. ~~The reason could come from the different ways they are obtained. It has been reported that taking an Arrhenius-type relationship between the strain rate and temperature can lead to an apparent activation energy value lower than the actual activation energy [18].~~

~~The energies obtained in both regimes can be used to extract the value of the nucleation and the migration energy separately, instead of only their sum.~~ It is important to emphasize that ~~this requires that~~ dislocation Burgers vectors and slip systems are the same in both regimes $A_1$ and $A_3$. This is fulfilled in the present case. However, the underlying idea in this work is the difficulty of dislocations to glide along <100> compared to the <110> direction. Therefore both in regimen $A_1$ and regimen $A_3$ dislocation glide is governed by the same process, i.e. the kink pair mechanism, but the higher kink pair nucleation rate and easier migration in regime $A_1$ obviously leads to a different dislocation microstructure in the two regimens. In the expression of the kink-pair nucleation energy, $F_k$' takes already into account the effect of a stacking-fault energy, which tends to increase the stress required to nucleate a kink pair. Then we can conclude that $F_k'(A_1) = F_k'(A_3) = F_k'$. However, there should be a difference in the kink-pair migration energy $Q'$, because if the kink-pair nucleation takes place ~~in the perfect dislocation, i.e.~~ correlatively in both partials, the situation is quite different than if it takes place only in one partial. This is due to the extra stacking-fault energy formed in the second case. If we consider $q(A_3) - q(A_1) = Q_D'(A_3) - Q_D'(A_1) = 0.5 \pm 0.2$ eV, this value should contain the difference in the migration energy. This extra energy for kink pairs to migrate should be proportional to the extra stacking-fault energy created when kink-pair nucleation takes place only in one partial. In order to calculate this extra energy we consider the extra stacking-fault energy originating from the migration of a kink across the secondary Peierls potential by a distance $a$ = 0.552 nm. With the periodicity of the primary Peierls potential, $h$ = 0.3905 nm and $\gamma$ = 340 ± 90 mJ/m$^2$ we obtain $Q_D'(A_3) - Q_D'(A_1) = \gamma \cdot h \cdot a = 0.45 \pm 0.15 \, eV$. As it can be observed this value fits quite well to the difference in the migration energy, supporting the idea that this difference comes from the extra stacking fault energy when a kink pair is expanding along only one partial. This difference could explain why in regime $A_1$ the dependence of the flow stress on the temperature is so flat and why in regime $A_3$ this dependence becomes so strong. And especially a change in the way that the kink pair nucleation takes place could explain the striking drop of the flow stress in regime $A_2$.

The lower value of the kink-pair migration energy in regime $A_1$ indicates that in this regime kink pairs expand quickly after formation. It is clear that a much lower $q$ value in regime $A_1$, where the long-segment limit is operating, is coherent with the quite weak dependence of the CRSS on temperature, indicating that dislocation glide is much easier than in the short-segment limit of regime $A_3$. Finally, we summarize by stating that the transition from the short- to the long-segment limit takes place in regime $A_2$ due to the vanishing of the preference for dislocations to glide along the <110> direction, which could be an explanation of the CRSS to drop in this regime.

The dislocation microstructure of strontium titanate was investigated at temperatures from 113 K to 296 K, in order to find an explanation for the sudden drop of the CRSS in regime $A_2$. Dislocations seem to have a preference to glide along the <110> direction at lowest temperatures. This tendency vanishes when temperature increases. Different kink-pair formation mechanisms were invoked, and a transition from the short-segment limit in regime $A_3$ to the long-segment limit in regime $A_1$ was found to justify this drop. Satisfactory fits were reached in each regime, and parameter values obtained from the best fit to experimental data have fully physical sense, supporting the idea that this transition could be the reason for the CRSS fall in regime $A_2$.

We are very grateful to Dr. D. Brunner for fruitful discussions and for providing the experimental data, to M. Kelsch and U. Salzberger for their help in TEM specimen preparation. This work has been supported by the German Research Foundation (DFG) through the project "Atomic-level theoretical and experimental study of lattice dislocations in perovskites".

**Figure 1.** Plot of the critical resolved shear stress of SrTiO$_3$ strained in regime A. The best fit in regimes A$_1$ and A$_3$ are shown.

**Figure 2.** Weak-beam TEM observation of SrTiO$_3$ strained at different temperatures; a) 113 K, b) 243 K, and c) 296 K taken from [4], showing how the preference of dislocation lines to be elongated along <110> directions is vanishing when temperature increases.

Figure1
**Click here to download high resolution image**

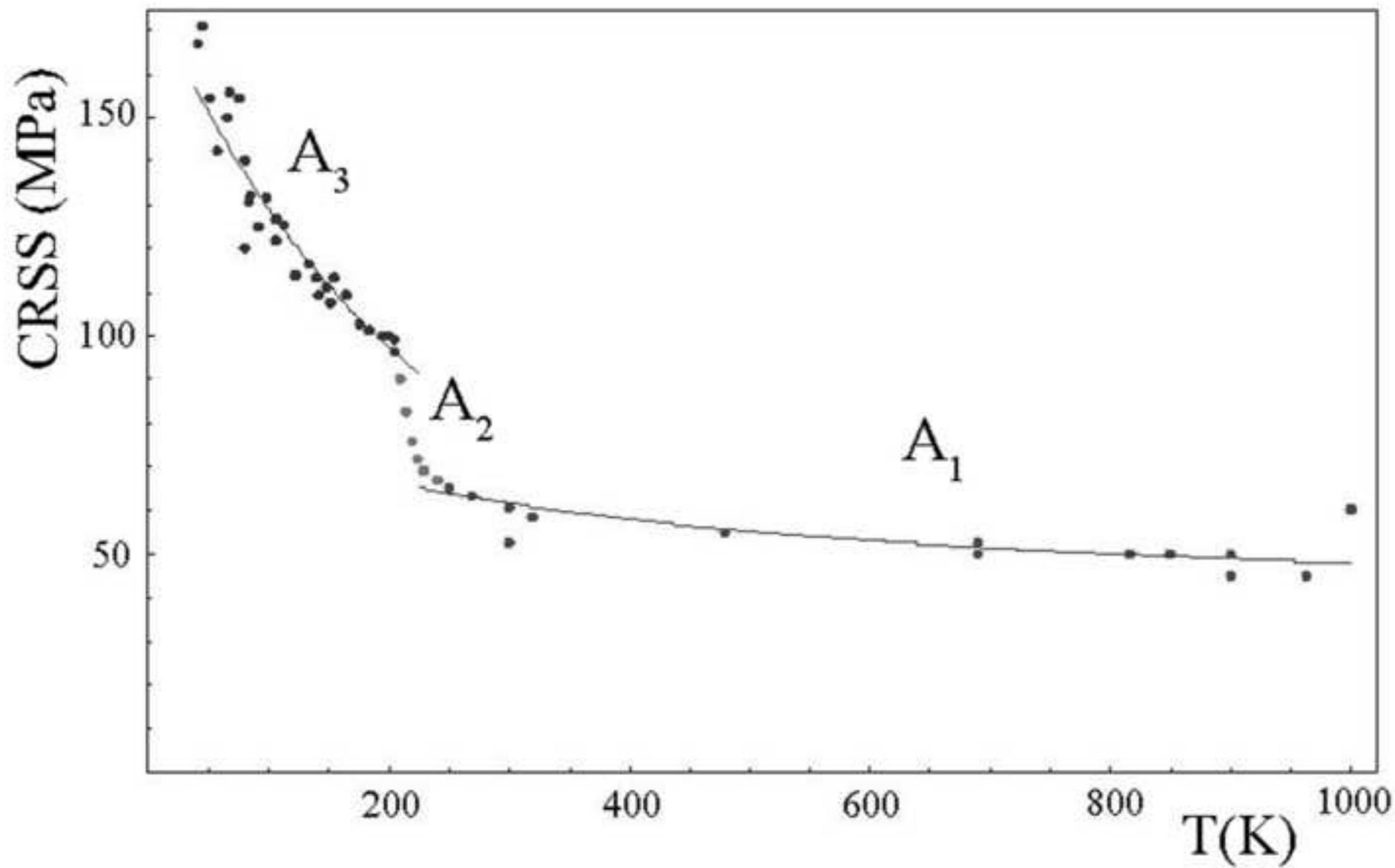

**Figure2**
**Click here to download high resolution image**

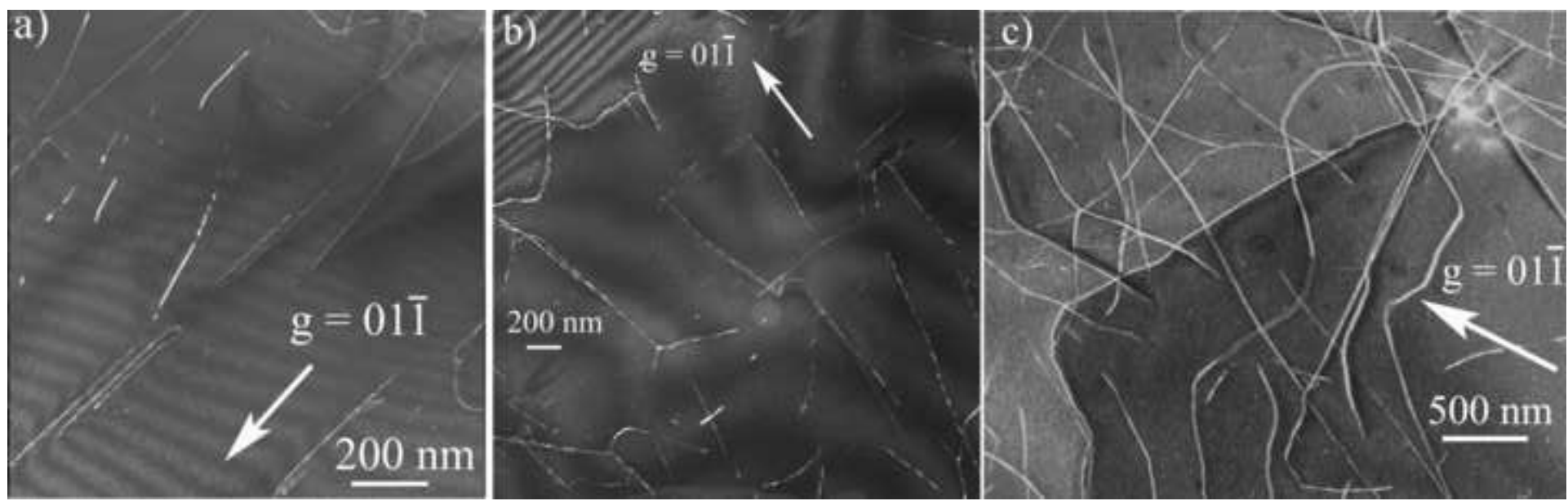